# Promoting Open Science Through Research Data Management

John A. Borghi, PhD*
Lane Medical Library
Stanford School of Medicine
Stanford University
Conceptualization, Writing, Original Draft, Writing - Review & Editing

Ana E. Van Gulick, PhD
Figshare
Conceptualization, Writing, Original Draft, Writing - Review & Editing

## Abstract

Data management, which encompasses activities and strategies related to the storage, organization, and description of data and other research materials, helps ensure the usability of datasets- both for the original research team and for others. For librarians and other service providers, describing data management as an integral part of a research process or workflow may help contextualize the importance of related resources, practices, and concepts for researchers who may be less familiar with them. Because data management practices and strategies overlap with those related to research reproducibility and open science, presenting these concepts together may help researchers realize their benefit at the laboratory level.

## Media Summary

The term "research data management" covers a range of activities related to how researchers save, organize, and describe the materials they work with over the course of a research project. Though often overlooked in formal coursework, data management is a necessary part of maintaining a record of what was done, when, and by whom. In this article, we discuss the connections between data management, reproducibility in science, and open science and provide some guidance for realizing the value of data management from the laboratory perspective.

# 1. What is Research Data Management?

Research data management includes a broad range of activities and strategies related to the storage, organization, and description of data and related research materials. Data management-related practices overlap substantially with those involved in data wrangling and data curation, which similarly involve rendering datasets into forms that ensure their usability by researchers and computational tools. But data management is broader, encompassing practices throughout the entire research lifecycle- from planning that occurs well before data is acquired through to the stewardship of data and other materials well after the conclusion of a research effort.

A typical workshop introducing data management to researchers may cover the importance of applying standardized file names and directory organizational schemes, best practices related to documentation, standards, and metadata, discussion of how to choose appropriate methods for backing up and archiving data, review of data management-related mandates and requirements (e.g. data management plans), and discussion of issues related to data security, privacy, licensing, and citation. Depending on the focus of the workshop, software-related practices such as version control, managing dependencies, and sharing code and computational environments may also be included. This broad range of topics reflects the importance of data management at two distinct but interrelated two levels:

1. For individuals and teams: Who need to keep track of data, materials, processes and procedures over the course of a given research effort.
2. For the broader research enterprise: That is invested in ensuring the integrity of the research process and reusability of datasets.

For individual researchers and teams, data management has an array of immediate and longer-term benefits. Good data management helps with quality control and prevents data from being lost or made inaccessible. Proper data management also increases the efficiency of the research process and collaborative work as every member of a research team is able to access and use the data, code, documentation, and other materials they need. For the broader research community, well-managed datasets can be more easily examined, (re)used, and built upon than those that have not been well organized or lack sufficient documentation. Because it is a necessary component of establishing what was done, when, and by whom, data management is also integral to establishing a record of the research process- a prerequisite for ensuring research integrity.

Data management is also, increasingly, a requirement. In 2023, the National Institutes of Health (NIH) will implement an updated Data Management and Sharing Policy (National Institutes of Health, 2020). Like similar policies implemented by other research funders including the National Science Foundation (NSF) and Patient-Centered Outcomes Research Institute (PCORI), this policy will require that researchers submit a data management plan- a short document outlining how they plan to manage their data and make it available to others- as part of any grant proposal. Other research data stakeholders, including scholarly publishers and research institutions, have also begun to implement policies related to data management. For example, the Stanford University data retention policy stipulates that Primary Investigators "Should adopt an orderly system of data organization and should communicate the chosen system to all members of a research group and to the appropriate administrative personnel" (Stanford University Research Policy Handbook, 1997).

Though the emphasis of the NIH policy on "maximizing the appropriate sharing of data" represents a significant step forward for the availability of data and other materials in the biomedical and health sciences, the policy landscape related to research data remains quite heterogeneous between and even within different data stakeholders (see Briney et al., 2015; Gaba et al., 2020). To encompass a wide range of practices and standards developed across the research community, implementation details are often left relatively general. However, because data management requires researchers to think prospectively about their practices, it provides an opportunity for researchers, librarians, policymakers, and other stakeholders to promote emerging scholarly communication practices such as the open sharing of articles, datasets, code, and other materials.

## 2. Data Management in Practice

Data management is an iterative and continuous process, related practices are implemented during the day-to-day course of a research effort and decisions made at early stages substantially affect what can be done later. A growing body of guidance and resources has been developed for researchers who need to manage the data and other materials associated with their work (e.g. Briney et al., 2020; Borghi et al., 2018; Broman and Woo, 2018; Goodman et al., 2014; Stoudt et al, 2021; Wilson et al., 2017). However, implementing proper data management is often quite complex in practice.

Consider as an example an experiment in which a human participant responds to a series of stimuli presented on a computer. Even this relatively straightforward setup presents substantial challenges for researchers trying to manage their data and other materials. Output files, containing response data for each participant, need to be cleaned and combined prior to analysis. Research software, which may include code created or adapted by the research team, is needed to present the stimuli, record participant responses, conduct data analyses, and create visualizations. Depending on the experiment, data and code may be accompanied by other research objects including consent forms, study stimuli, or paper surveys or questionnaires. Making use of all this requires documentation which, at the very least, describes data collection and analysis procedures (i.e. protocols), the contents of data files (i.e. data dictionaries), and details of the computational environment needed to run related code and software.

For the research team, all of these objects - datasets, code, research materials, documentation, etc - need to be properly managed even if there is no intention of sharing them outside of the research group. Doing so properly requires extensive planning and, as summarized in Table 1., consideration of manifold issues and activities. Despite the importance of data management, related practices and strategies are often not covered in coursework (Tenopir et al., 2016) and are instead learned informally from peers (Borghi and Van Gulick, 2018; 2021). As a result, even when individual researchers exercise relatively good data management, their practices may not be standardized between projects or within their own research group.

|  | **Project Planning** | **Data Collection and Analysis** | **Data Publication and Sharing** |
|---|---|---|---|
| Data Management Activities | - Data Management Planning | - Saving and backing up files<br>- Organizing files<br>- Formatting and describing data according to standards<br>- Maintaining documentation and metadata | - Preserving data and other materials (e.g. reagents, code)<br>- Assigning persistent identifiers |
| Open Science Activities | - Planning for open (e.g. including data sharing in consent forms) | - Using open source tools<br>- Using transparent methods and protocols | - Sharing data<br>- Publishing research reports openly (e.g. open access publishing). |
| Reproducibility-Related Activities | - Preregistering study aims and methods<br>- Using appropriate research designs (including sufficient statistical power) | - Preventing methodological issues (e.g. p-hacking, HARKing)<br>- Implementing quality control measures | - Preventing publication bias<br>- Following reporting guidelines |

**Table 1.** Activities related to data management, reproducibility, and open science at different stages of a research effort.

The terms describing different stages are defined quite broadly. Project planning encompasses both the development of research questions and practical steps. Data collection and analysis could involve researchers generating data or acquiring data initially collected by others. Data publication and sharing may involve describing results in a manuscript or sharing datasets through a repository. This list is not exhaustive but is intended to demonstrate that activities related to data management, reproducibility, and open science are contiguous.

Academic libraries have positioned themselves as a source of guidance for researchers on topics related to data management (Cox et al., 2019; Tang and Hu, 2019; Tenopir et al., 2014) as have other stakeholders including scholarly publishers, funding agencies, and data repositories. However, librarians, researchers, and other data stakeholders have different perspectives and incentives, which complicates communication about standards and best practices. For example, a librarian may recommend that a researcher deposit data and code in repositories with robust preservation strategies where materials are described with rich metadata and assigned persistent identifiers to help ensure discoverability and citability. A researcher who typically works with data they have acquired themselves and is primarily incentivized to publish high-impact papers may lack the context necessary to appreciate the importance of such practices or see them as unnecessary extra steps.

One entry point to overcoming differences in perspective when discussing data management is by focusing on research processes or workflows. In this context, a workflow is defined as the series of programmatic steps or practical "ways of doing things" as data is collected, processed, and analyzed. As shown in Figure 1., there are multiple points in any research effort where researchers make decisions that can significantly affect their results. A research workflow specifies the steps a research team actually takes among many possibilities and directions not taken. Though the term most often refers to data cleaning and analysis procedures, a well-defined workflow may also include standard operating procedures for organizing files, preserving backups, maintaining documentation (e.g. protocols, data dictionaries) and research-related code, as well as guidelines for how the data should be made available to others (e.g. how it should be formatted, described, and organized, what repositories should be used).

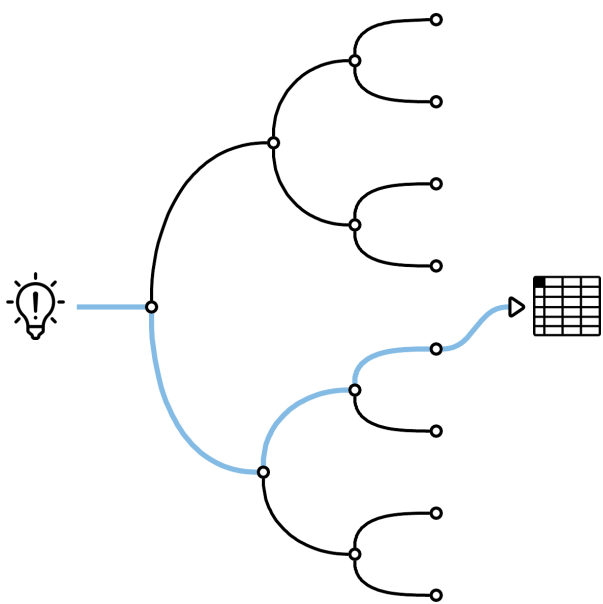

**Figure 1:** A highly simplified model of the research process, starting with the generation of a research question through to the communication of research results in a publication or shared dataset.

The highlighted line represents the programmatic steps taken in a given project, while the black lines represent directions not taken. This model is highly iterative, decisions made at early stages affect subsequent decisions and outcomes. A research workflow includes methodological details (e.g. data collection parameters, data analysis procedures) as well as more practical processes such as how data is to be stored, organized, and described. For computationally intensive research, data management may consist of activities such as version control and the maintenance and preservation of computational environments.

Using research workflows to frame data management puts related practices into a context familiar to researchers. Ideally, much of a workflow is planned in advance and builds towards fulfilling the goals of the research effort. Incorporating data management into research workflows may help related practices be seen as part of the day-to-day task of doing research rather than "extra work".

For librarians and other stakeholders, conceptualizing the research workflow as the locus of support can help reinforce that research data management is about data and other materials that are *in process*, meaning that they are acquired, processed, analyzed, archived, and shared in order to be acted upon in some way. "Best practices" from the data stewardship perspective need to be balanced with a constellation of other motivations, incentives, and needs, such that the resulting "good enough" practices enable the research team to complete all of their goals, including the curation and dissemination of accessible and (re)usable datasets.

## 3. Realizing the Benefits of Data Management

Critical to the realization of the benefits of data management, both for a research team and for the research enterprise more generally, is the development and adoption of standards. A standard specifies how exactly data and related materials should be stored, organized, and described. In the context of research data management, the term typically refers to the use of specific and well-defined formats, schemas, vocabularies, and ontologies in the description and organization of data. However, for researchers within a community where more formal standards have not been well established, it can also be interpreted more broadly to refer to the adoption of the same (or similar) data management-related activities or strategies by different researchers and across different projects.

Formal data standards are developed and maintained by an array of data stakeholders, including the research community itself. For example, the neuroimaging community developed the Brain Imaging Data Structure (BIDS) (Gorgolewski et al. 2016; RRID:SCR_016124) to standardize the description and organization of raw MRI data. BIDS was subsequently extended to other imaging modalities (i.e. EEG, MEG, PET) and is now integral to an ecosystem of quality assurance, processing, and analysis tools (Gorgolewski et al., 2017). The International Neuroinformatics Coordinating Facility (INCF), a standards organization dedicated to open and FAIR neuroscience (Abrams et al., 2021), provides support for BIDS and formally endorsed it as a standard in 2019.

The use of common data elements (CDEs), wherein well-defined questions (variables) are paired with a discrete set of allowable responses (values) that are used in standardized ways across different research efforts, is particularly relevant in light of the new NIH Data Management and Sharing Policy. The promise of CDEs is that their use can facilitate comparisons between studies and simplify data aggregation and meta-analysis (Sheehan et al., 2016). To this end, NIH maintains a CDE repository (https://cde.nlm.nih.gov/) to provide access to structured definitions of the data elements recommended by its institutes and centers as well as other organizations.

Though they are not strictly standards, the FAIR Guiding Principles (Mons et al., 2017; Wilkinson et al., 2016) also provide a starting point for working through considerations related to data management and sharing. Because the FAIR Principles were developed to describe the characteristics that data-related infrastructure should adopt to facilitate data reuse, Table 2. outlines how the principles of <u>F</u>indable, <u>A</u>ccessible, <u>I</u>nteroperable, and <u>R</u>eusable can inform data management practices implemented within a given research effort.

Data repositories play a key role in maintaining and promoting standards. BIDS is central to the OpenNeuro Repository (Markiewicz et al., 2021), which facilitates the sharing and reuse of neuroimaging data. Similarly, the Inter-university Consortium for Political and Social Research (ICPSR), promotes the Data Documentation Initiative (DDI) (Vardigan et al., 2008) as a standard for survey data. Figure 2. outlines information for identifying an appropriate repository to deposit a given dataset. A similar representation, which emphasizes the use resources developed and implemented by relevant research communities, could also be used for identifying which data standards should be used to structure and describe a given dataset.

**Table 2:** Extension of the FAIR guiding principles

The FAIR Principles were initially developed to apply to data-related infrastructure and emphasize machine actionability. However, they imply a number of data management-related activities and strategies that can be implemented by researchers in their day-to-day work. Additional information about choosing the appropriate repository for a dataset can be found in Figure 2.

| FAIR Principle | For Infrastructure | For Researchers |
|---|---|---|
| **Findable** | Data and metadata should be easy to find by both humans and computers.<br><br>In practice, this means that data should be assigned unique and persistent identifiers, described using rich metadata, and registered or indexed in a searchable resource. | Research teams should implement standardized practices related to organizing files (e.g. standardized file naming conventions) so data can be found when needed.<br>When data is made available to others, it should- whenever possible- be uploaded to a repository that assigns a persistent identifier (e.g. DOIs, RRIDs, etc) and describes datasets with standardized metadata.<br><br>Complete and high quality metadata should be added so data can be discovered and linked to related resources (e.g. related paper DOIs, author ORCIDs). |
| **Accessible** | There is a clearly defined method for accessing the data. Data should be retrievable by its identifier using a standardized protocol that is open, free, and universally implementable. Metadata should be accessible even when data is no longer available. | Data is available through a clearly defined process. Members of the research team should be able to access raw data, intermediate products, and other research materials.<br><br>When data and other materials are made available to others, there should be a clear path to gaining access. The terms by which the data will be made available (e.g. to whom, when, and for what purpose) should be articulated and abided by. |
| **Interoperable** | Data should be usable across a range of applications and workflows. Data should use formal, accessible, shared, and broadly applicable models for knowledge representation. | Data should be structured in a standard way so it can be easily combined with other similarly structured datasets. In practice, this means implementing a range of practices such as describing and organizing data (e.g. applying appropriate metadata, maintaining data dictionaries) and saving files in open or non-proprietary file formats. |
| **Reusable** | Metadata and data should be described following relevant community standards and have clearly defined conditions for reuse including a machine-readable license | Data should be saved, organized, and described with its future (re)use in mind. A future user may be a member of the research team who is returning to the data after a period of several months or years or another researcher who is (re)using the data for another purpose. |

Resources like FAIRsharing.org (https://fairsharing.org/) and the Registry of Research Data Repositories (https://www.re3data.org/) provide information related to the standards and repositories for specific data types and disciplinary communities. Similarly, the forthcoming NIH data policy includes a set of "desirable characteristics for all research repositories" (i.e. assigns persistent identifiers, has a plan for long-term sustainability) to inform decisions about platforms for managing and sharing data resulting from federally funded research.

In research areas for which standards or best practices have not been developed or widely adopted, discussion of data management may be relatively nascent. Even with guidance from a data librarian or other data management experts, developing a research team's data management practices can be an overwhelming experience, but could begin with a conversation catalyzed by questions such as:

1. Would every member of the research team be able to find and use the data, code, documentation, and other materials related to this project?
2. Would another researcher who works in the same field be able to find and use the data, code, documentation, and other materials related to this project?
3. Ten years from now, would you or another researcher be able to find and use the data, code, documentation, and other materials related to this project?

These questions are drawn from a longer checklist which is attached in full as a supplementary document. The complete checklist was initially developed to accompany one-time data management workshops attended by researchers. The intent was for workshop attendees to use questions such as these as a starting point for discussing data management-related practices with their research groups. This approach has subsequently formed the basis of focus groups and survey instruments examining data management-related practices, perceptions, and needs within Stanford University's School of Medicine. Answering in the affirmative to all three questions or even all he items on the checklist does not indicate that a given researcher or research group is engaged in optimal data management. Instead, the questions provide an opportunity to consider gaps in current practices.

## 4. Data Management is Necessary for Reproducibility

For both individual research teams and the research enterprise more broadly, one of the most cited benefits of good data management is the foundational role related practices play in establishing reproducibility. Efforts to address reproducibility are generally concerned with establishing the credibility, reliability, and validity of scientific research (Goodman et al., 2016; Label et al., 2018; Peng and Hicks, 2021; Peterson and Panofsky, 2021; Plesser, 2017; see also Devezer et al., 2021) and, as demonstrated in Table 1., address a wide array of issues related to study design, analytical practices, and the communication of research results.

Methods for addressing reproducibility-related concerns often center on researchers communicating the details of their research process. This includes both enhanced transparency for study-related decisions and procedures and appropriate sharing of research materials, including datasets. Lack of transparency has substantial consequences for reproducibility. The effect of the phenomenon shown in Figure 1., commonly described as "researcher degrees of freedom" (Simmons et al., 2011) in reproducibility-related literature, has been demonstrated empirically by initiatives where multiple research teams independently examine the same dataset (e.g. Botvinik-nezer et al. 2020; Silberzahn et al. 2018). In the absence of extensive documentation, teams of experts may construct different workflows that lead to different results from the same data. This underlines the necessity of being able to trace the exact process used to get to a set of research results.

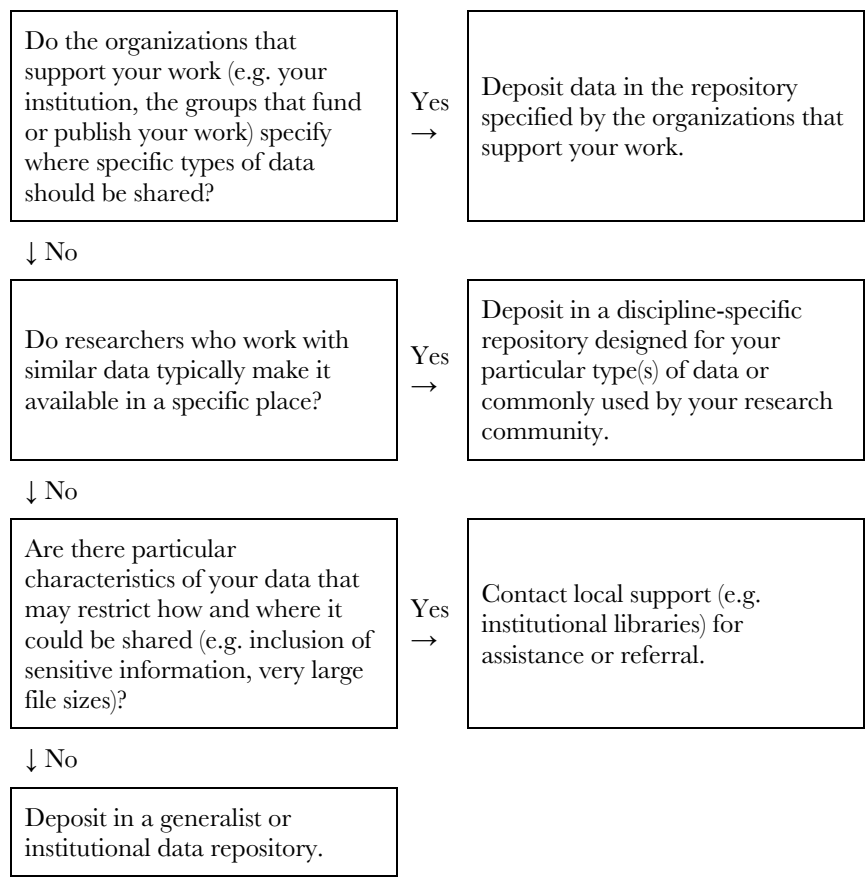

**Figure 2:** Guidance for researchers when choosing a data repository

The different categories of research repository implied in this figure (institutional, generalist, specialist) are not necessarily mutually exclusive. For example, a funder or publisher may specify that data should be deposited in a specific generalist repository (e.g. Dryad, Zenodo, Figshare).

In general, researchers should strive to deposit their data in a repository where it will be easily found by others.

While a full accounting of every practice proposed to address reproducibility over the course of a research effort is beyond the scope of this manuscript (see National Academies of Sciences, Engineering, and Medicine, 2019), many of these practices overlap with those that fall under the umbrella of research data management and similarly involve careful consideration of issues throughout the research process. For example, instruments such as the Materials Design Analysis Reporting (MDAR) Checklist (Macleod et al., 2021) are intended to address reproducibility-related concerns by enabling unambiguous descriptions of what research materials were used (e.g. reagents, model organisms) and how data, code, and other elements of the research process can be accessed. While such checklists are typically completed at the time of publication, completing them requires researchers to have carefully documented their processes and materials throughout the research process.

Beyond checklists and other individual interventions, proper data management is essential to establishing an audit trail or record of the research process. In practice, many of the same activities and strategies that help researchers keep track of their data, code, materials, practices, and procedures, during the day to day course of working on a research project also help to ensure that their processes and conclusions are reproducible.

## 5. Data Management is Integral to Open Science

For an individual research team, it may not be completely clear how to establish reproducibility, even internally. One starting point would be to pursue a narrow definition of the term, where a research effort is said to be reproducible if the same results are found when the same analytical pipeline is applied to the same data. Often called computational or analytic reproducibility (Label et al., 2018; Stodden et al. 2018), this can still be difficult in practice. In computationally intensive research, changes in software version and operating system can have measurable effects on research results (e.g. Gronenschild et al., 2012). Therefore achieving computational reproducibility may require not just careful organization and documentation but also the application of tools such as software containers that include the code needed to reproduce the analysis as well as the specific software version and operating system used (Grüning et al., 2018).

Establishing reproducibility within the research enterprise more broadly requires the adoption of a range of practices, many of which fall under the umbrella of open science (Munafo et al., 2017; National Academies of Sciences, Engineering, and Medicine. 2018; 2019). The term Open Science covers a variety of efforts focused on making scientific research more transparent and accessible. Though it is frequently used to refer to efforts aimed at ensuring access to tools and research products, open science also encompasses efforts to ensure that the scientific enterprise is inclusive and equitable. Such efforts are interrelated but, for the purposes of this review, we are focused primarily on openness for research objects (e.g. software, publications, protocols, datasets) (see Table 1). Related practices exist along a continuum, meaning that it is generally more accurate to describe a research effort's degree of openness rather than categorizing it as simply "open" or "closed".

As of this writing, perhaps the clearest demonstration of the immense value of open science is the publication of the complete Sars-CoV-2 genome, first on the Virological discussion forum (https://virological.org/t/novel-2019-coronavirus-genome) and subsequently on Genbank (Wu et al., 2020), which catalyzed efforts to create tests and interventions targeting the disease. Moderna's COVID-19 vaccine, based on this genome, was first sequenced just days after the initial posting (Moderna Therapeutics, 2021).

For individual researchers and teams, adopting open science practices has a number of potential benefits, including exposure to new tools and methods and streamlining the research process (Allen and Mehler, 2019; Lowndes et al., 2017). Returning to the example of an experiment where human participants respond to stimuli on a computer screen, the research team may develop open source tools to collect and analyze their data, may publish detailed information about their protocols and analytical pipelines so they can be verified and built upon by others, may post a preprint to quickly disseminate their findings, may disseminate their work broadly through using one of several routes to open access, or may make data, code, stimuli, or other materials available through an open repository. Each of these practices have benefits for the research enterprise- open science can quicken the pace of research and the reuse of shared materials has innumerable downstream benefits. However, depending on their needs and priorities, such benefits may not be immediately clear or compelling for the research team putting conducting open science.

Despite the benefits, open science can be difficult to put into practice. Implementing all the practices outlined in Table 1. effectively requires planning. For example, sharing data from human participants requires consideration of how to handle personally-identifying information throughout the entire research process. Depending on the nature of the data, deidentification or anonymization may not be possible (e.g. Rocher et al. 2019) meaning that, while it still may be possible to provide access to the data to certain individuals under certain circumstances, it can not be shared publicly.

As with data management, open data sharing requires consideration of issues that may be outside of a research team's expertise. There may or may not be discipline or data-type specific standards for exactly what data should be openly shared (i.e. raw data vs. processed data), what format it should be shared in, or how it should be licensed and made available. There is also a difference between data that is openly available and data that has been shared in a truly usable form. Absent guidance from data management experts, interventions targeted at making data and other materials available may not necessarily result in data and other materials being shared in a (re)usable form. Datasets may need to be made available alongside code, explanations, and other elements of the research process to be reproducible (Chen et al., 2019).

There has been extensive research into why researchers do and do not share their data openly, and the major themes that arise are lack of time and skills necessary to organize the data into a form suitable for sharing (see Perrier et al. 2020). This supports the notion that, for many research communities, data sharing and other open science-related activities may not be part of a research team's regular workflow and be seen as "extra work" that is not rewarded by institutions or funders or in the hiring or promotion process.

For researchers, librarians, and other data stakeholders, data management practices provide an entry point for promoting open science practices like data sharing. Data management can be positioned as a solution to immediate needs, such as ensuring that data is accessible and secure. But the same practices that help ensure data and other materials are usable by the research team as they are working with them also make the process of sharing them openly substantially more efficient. For example, outlining methods for maintaining internal documentation about study protocols provides an opportunity to promote sharing them through a tool like Protocols.io (https://www.protocols.io/), discussing solutions for the long term preservation of data and code provides an opportunity to promote publishing them in discipline-specific or generalist repositories and providing guidance on how data and other research products should be cited in published literature provides an opportunity to promote persistent identifiers (e.g. DOIs) as well as the posting of preprints and open access to research articles.

## 6. Data Management is Good Research Practice

Data management, reproducibility, and open science are interrelated. Any one of them can be an angle for promoting the others based on a research team's existing workflow, priorities, and needs. On their own, implementing good data management practices is not sufficient to establish reproducibility. However, a set of research results cannot be efficiently examined or replicated if the underlying data, code, and other materials were not properly saved and organized and analysis-related procedures and decisions were incompletely documented. For a given experiment, it is also unlikely that all of the data, code, documentation, and other materials need to be shared. Depending on the nature of the research effort and the conclusions being drawn, it is possible that only the raw data, only the "final" fully processed dataset, or a selection of intermediate data products need to be shared to establish reproducibility. However, to establish a record of what was done, when, why, and by whom all of it needs to be well managed and stewarded throughout the research process.

On their own, policies related to data management and data sharing have had mixed success in ensuring data is made available in a usable form (Couture et al., 2018; Federer et al., 2018; Parham et al., 2016). Even when data and other materials are ostensibly available, they may not in fact be "available upon request" when requested (Vines et al., 2014; Stodden et al. 2018), not actually deposited in a repository (Danchev et al., 2021; Van Tuyl and Whitmire, 2016), or not shared in a usable or reproducible form (Hardwicke et al., 2018; 2021). Similarly, when presented in isolation, activities related to data management, reproducibility, and open science may not resonate with researchers who have different motivations and incentives.

Reproducibility and open science begin in the laboratory, with practices implemented by researchers during the day-to-day course of a research effort. Visualizations such as the research data lifecycle (Ball et al., 2012; Griffin et al., 2018) are often used to describe breadth of data management activities or strategies, but situating data management instead as an integral part of a research workflow may help contextualize such practices and prevent them from being seen as just extra work to be done at a discrete point in the research process. Grounding such practices in a context that researchers are familiar with also provides an opportunity for promoting other practices that may involve issues that a research team may not be familiar with, including automated workflows, large-scale data reuse, and open source research infrastructure.

## Disclosure Statement

Both authors work broadly in the field of data management and sharing. A.E.V. is currently employed by the commercial company, Figshare. Support from this employer was provided in the form of the author's salary, but the employer has not influenced the development or content of this project nor the decision to publish this work.

## Supplementary Files

The included data management checklist is intended as a starting point for groups looking to realize the benefits of research data management from the laboratory side. The guide is designed to be easily customizable and extensible, so it is likely that practices specific to particular research communities or data types are not included. Because of their close relationship with data management, going through this checklist also provides an opportunity to promote activities related to establishing reproducibility and open science.

# Data Management Checklist

**How to use this checklist:**

This checklist is intended to help you get started integrating data management into your research practice. This checklist can help you identify gaps and communicate elements of data management to members of your research team. It is recommended that you apply this checklist to an individual research project as practices and procedures may vary considerably between projects.

The focus of this checklist is on practices implemented by you and your collaborators. When managing research data, it is extremely valuable to seek out and follow community standards and practices rather than developing practices from scratch. For assistance identifying standards or practices that may have been developed for your research area, contact [organization-specific or community-specific data management expertise].

This checklist is not intended to cover every single aspect of research data management. For example, the current iteration does not cover practices or strategies related to quality control in depth. It is likely that certain data management practices that are specific to your research, your type(s) of data, and your needs as a researcher are not covered. It is also possible that certain items on the checklist will not apply to the specific type(s) of data you are working with.

Research data management exists along a continuum and following best practices often involves adopting new guidance, procedures, and techniques as they are developed. So it is worth revisiting this checklist with project team members periodically.

Please feel free to modify this checklist or adapt it to better fit your needs.

If you have any questions about this checklist or research data management more generally, contact [organization-specific or community-specific data management expertise].

**Project Name:**
**Project Team Members:**
**Last Date Reviewed:**

---

If you and your research team can confidently check off the following three prompts for your project, you are probably doing a reasonable job of managing your data. Because these prompts are so broad, we would encourage you to discuss *how* you are implementing related practices and strategies and working through the rest of the checklist with your team.

- Every member of the research team is able to find and use the data, code, documentation and other materials related to this project.
- Another researcher who works in the same field would be able to find and use the data, code, documentation and other materials related to this project.
- We believe we will be able to find and use the data, code, documentation, and other materials related to this project ten years from now.

The following items relate to specific data-related practices. If you and your research team can confidently check off the following prompts, you have started the process of integrating good data management practices into your project. In this context, standardized practices are those that are consistently employed by every member of the research team.

*We have done the following at the beginning of our project:*

- We have reviewed all applicable policies related to our data (including from our institution, funder, potential publisher, etc)
- We have read through and understand other relevant agreements, licenses, or other requirements related to our data (e.g. data use agreements, IRB or funder policies).
- We have sought out community standards and best practices related to our data.
- We have discussed the intended products of this project (papers, datasets, software tools, etc) and have decided to what extent we will be making our data and other materials available to others.

*We have a plan:*

- We maintain documentation that describes the type(s) of data we are collecting/analyzing/working with over the course of the project as well as details about materials that are needed to understand and use the data (documentation, code, etc).
- We maintain documentation that describes the specific data management practices (e.g. file naming, formats, and standards, backing up data) employed throughout the course of this project.
- We maintain documentation that outlines the roles and responsibilities of individual team members related to managing data (e.g. maintaining good documentation, following file naming conventions, etc) as well as who is ultimately responsible for ensuring the data is properly managed throughout the course of the project and following its conclusion.
- Members of the research team have access to the above documentation and review it periodically.

*We are keeping our data organized:*

> We have a standardized set of practices related to saving datasets and other project materials while we are working with them (e.g. digital data is saved on a lab server).
> Our practices related to saving data are in line with [our organization's] risk classification system and, when possible and appropriate, include multiple backups.
> We have standardized conventions for naming project-related objects and files (including datasets) that enable us to quickly identify the materials we are looking for.
> We have standardized systems for organizing project-related objects or files that enable us to easily find the materials we are looking for (e.g. a standardized file structure).
> When applicable, we have standardized systems for naming and organizing information within our data files (e.g. standardized variable names, tidy spreadsheets).
> Our practices related to saving, organizing, and describing data files have been optimized to facilitate quality control.
> Our practices related to saving, organizing, and describing data files are in line with community standards and best practices.

*We are keeping good records:*

> We maintain documentation that describes how we keep datasets and other materials organized while we are working with them (e.g. naming conventions, file structures, etc).
> We have standardized procedures for documenting the structure and contents of individual data files (e.g. maintaining codebooks, data dictionaries, etc).
> We have standardized procedures for documenting project-related decisions, steps, procedures, and workflows (e.g. maintaining protocols, lab notebooks, etc).
> We have standardized procedures for saving and versioning research-related code and other elements of the research process (e.g. workflows, software containers).

*We have done (or will do) the following before the end of the project:*

> When necessary and appropriate, datasets and related materials are converted into a form suitable for long-term storage or archiving (e.g. open file formats for digital files).
> If we have decided to share our data, we have uploaded it to a suitable data repository to make it discoverable and FAIR together with any documentation and materials that are necessary to make use of it.
> We have moved project-related data, documentation, and other materials to a location suitable for long-term storage or archiving that we are able to access when necessary.

*We are checking up on ourselves:*

> Members of the research team are actually following the practices and procedures we have decided upon.
> Study documentation is updated regularly to reflect any changes to data management-related practices and procedures.
> We have adopted community standards and practices whenever possible,
> We have established procedures for onboarding new team members about our data management practices, educating members about changes to existing practices, and managing data as team members move onto new projects.